\documentclass[aps,prl,twocolumn,amssymb,amsmath]{revtex4}
\usepackage{graphicx,latexsym,pifont,dsfont}
\usepackage{tikz}
\usetikzlibrary{snakes}
\definecolor{labelkey}{cmyk}{.4,.2,0,0}
\usepackage{color}
\definecolor{Blue}{rgb}{0.00, 0.00, 1.00}
\definecolor{Red}{rgb}{1.00, 0.00, 0.00}

\newcommand{\rme}{{\mathrm{e}}}
\newcommand{\rmd}{{\mathrm{d}}}

\newcommand{\half}{\frac12}

\newcommand{\Fig}[1]{\includegraphics[width=8.7cm]{./#1}}

\renewcommand{\epsilon}{\varepsilon}

\def\be{\begin{equation}}
\def\ee{\end{equation}}

\def\bal{\begin{align}}
\def\eal{\end{align}}

\def\bea{\begin{eqnarray}}
\def\eea{\end{eqnarray}}
\renewcommand{\log}{\ln }

\usepackage{color}

\begin{document}

\title{The Maximum of a Fractional Brownian Motion: Analytic Results from Perturbation Theory}

\author{Mathieu Delorme and Kay J\"org Wiese}
\address{CNRS-Laboratoire de Physique Th\'eorique de l'Ecole
Normale Sup\'erieure, 24 rue Lhomond, 75005 Paris, France
}

\begin{abstract}
\medskip
Fractional Brownian motion is a non-Markovian Gaussian process $X_t$, indexed by the  Hurst exponent $H$. It  generalises standard Brownian motion (corresponding to  $H=1/2$). We study the probability distribution of the maximum $m$  of the process and the time $t_{\rm max}$ at which the maximum is reached. They are encoded in a path integral, which we evaluate perturbatively around a Brownian,  setting $H=1/2 + \varepsilon$. This allows us to derive analytic results beyond the scaling exponents. Extensive numerical simulations for different values of $H$  test these analytical predictions and show excellent agreement, even for large $\varepsilon$.
\end{abstract}

\maketitle

Random processes are ubiquitous in nature. While  averaged quantities have been studied extensively and are well characterized, it is often more important to understand the extremal behavior of these processes \cite{GumbelBook}, associated with failure in fracture or earthquakes, a crash in the stock market, the breakage of dams, etc. 

Though many processes can successfully be modeled by Markov chains and are well analyzed by   tools of statistical mechanics, there are also interesting and realistic systems which do not evolve with independent increments, and thus are non-Markovian, \textit{i.e.}\ history dependent. Dropping the Markov property, but demanding that a continuous process be scale-invariant and Gaussian with stationary increments  defines an enlarged class of random processes, known as fractional Brownian motion (fBm). Such processes appear in a broad range of contexts: Anomalous diffusion \cite{BouchaudGeorges1990}, polymer translocation through a pore \cite{ZoiaRossoMajumdar2009,DubbeldamRostiashvili2011,PalyulinAlaNissilaMetzler2014}, the dynamics of a tagged monomer \cite{GuptaRossoTexier2013,Panja2011}, finance (fractional Black-Scholes and fractional stochastic volatility models \cite{CutlandKoppWillinger1995}), hydrology  \cite{MandelbrotWallis1968},  and many more. 

\begin{figure}[t]
	\Fig{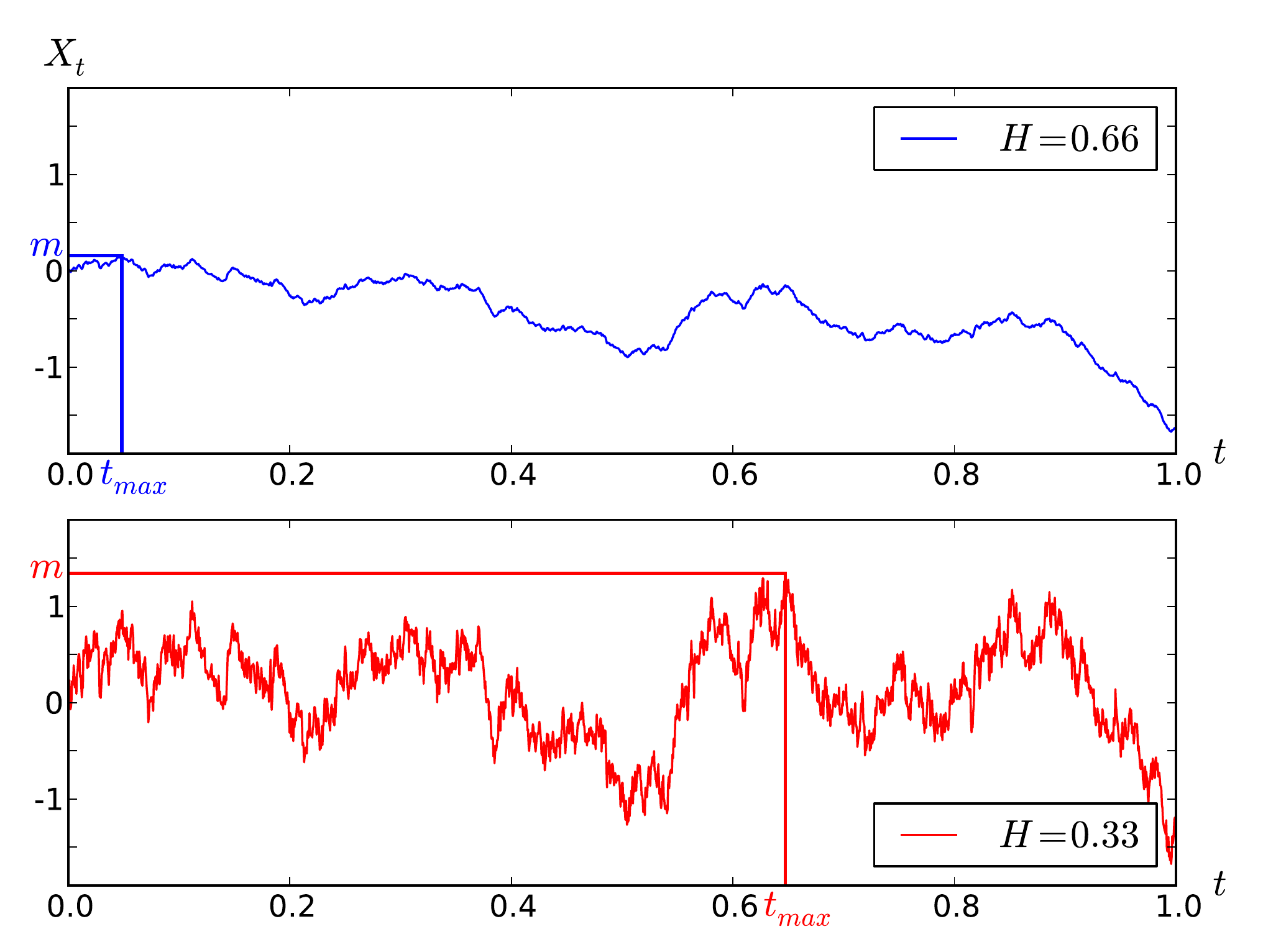}
	\caption{Two realisations of fBm paths for different values of $H$, generated using the same random numbers for the Fourier modes in the    Davis and Harte  procedure  \cite{Dieker}. The observables $m$ and $t_{\rm max}$ are given.}
	\label{Illustration}
\end{figure}

FBm is a generalization of  standard Brownian motion to other fractal dimensions, introduced in its final form by Mandelbrot and Van Ness \cite{MandelbrotVanNess1968}. It is a  Gaussian process $(X_t)_{t\in \mathbb{R}}$, starting at zero, $X_{0}=0$, with  mean $\left<X_{t}\right>=0$  and  covariance function (variance)
\begin{equation}
\langle X_t X_s \rangle = s^{2H} + t^{2H} - |t-s|^{2H} \ .
\end{equation}
The parameter $H\in(0,1)$ is the Hurst exponent; the process typically grows with time as $t^H$. 
Standard Brownian motion corresponds to $H=1/2$; there  the covariance function reduces to $\langle X_t X_s \rangle = 2 \min(s,t) $. 
Unless $H=1/2$, the process is non-Markovian, \textit{i.e.}\ its increments are not independent: For $H>1/2$ they are correlated, whereas for $H<1/2$ they are anti-correlated, 
\be
\langle \partial_{t} X_t \,\partial_{s} X_s \rangle = 2H (2H-1) |s-t|^{2(H-1)} \ .
\ee 
In this letter we study the {\em maximum} of a fractional Brownian motion $m= \max_{t\in[0,T]}X_t$ and the time $t_{\rm max}$ when this maximum is reached  \footnote{These two random variables are almost surely well defined, as realisations of $X_t$ are continuous and realizations where the maximum is degenerate are of  measure zero.} with the initial condition $X_0 =0$ and total time  $T>0$. Figure \ref{Illustration} shows an illustration for different values of $H$, using the same random numbers for the Fourier modes. We will denote $P^T_H(m)$ and $P^T_H(t)$ their respective probability distributions. Previous studies  can be found in \cite{Sinai1997,Molchan1999}.

These observables are closely linked to other quantities of interest, such as the first-return time, the survival probability, the persistence exponent, and the statistics of records. Though studied since long time, most results for  non-Markovian processes are quite recent \cite{DerridaHakimZeitak1996,Majumdar1999,MajumdarRossoZoia2010}.

Following the ideas of \cite{MajumdarSire1996,OerdingCornellBray1997,WieseMajumdarRosso2010}, we  encode our observables  $P_H^T(m)$ and $P_H^T(t)$ in a path-integral, \begin{equation}
\begin{split}
 Z^+&(m_1,t_1;x_0;m_2,t_2) = \\
&\int_{X_0 =m_1}^{X_{t_1+t_2}=m_2} \mathcal{D}[X] \,\theta[X] \, \delta(X_{t_1} - x_0)\, e^{-S[X]}\ .
\label{PathIntegral}
\end{split}
\end{equation}
This   sums over all paths $X_t$, weighted by their probability $e^{-S[X]}$,  starting at $X_0=m_1>0$ (shifted for convenience), passing through $x_0$ (close to $0$) at time $t_1$, and ending in $X_{t_1+t_2}=m_2>0$, while staying {\em positive} for all $t\in \left[0,t_1+t_2 \right]$.  The latter is enforced by the product of  Heaviside functions $\theta[X]:= \prod_{s=0}^{t_1+t_2}\theta(X_s)$.\\

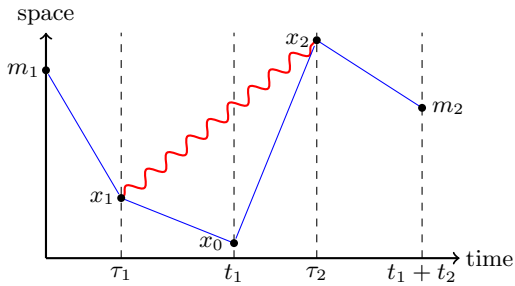
\begin{figure}
        {{\begin{tikzpicture}
                                                \draw [->,thick] (0,0) -- (5.5,0);
                        \draw [->,thick] (0,0) -- (0,3);
                        \draw [dashed] (2.5,0) -- (2.5,3);
                        \draw [dashed] (5,0) -- (5,3);
                        \draw [dashed] (1,0) -- (1,3);
                        \draw [dashed] (3.6,0) -- (3.6,3);
                                                \node (m1) at  (0,2.5) {$\hspace{-6mm}m_1$};
                        \node (x1) at (1,0.8) {$\hspace{-5mm}x_1$};
                        \node (x2) at (3.6,2.9) {$\hspace{-5mm}x_2$};
                        \node (x0) at (2.5,0.2) {$\hspace{-6mm}x_0$};
                        \node (m2) at  (5,2) {$\hspace{-6mm}\parbox{0mm}{$~~~~~~m_2$}$};
                                                \draw [snake=snake,red,thick] (x1) -- (x2);
                                                \draw [blue] (m1) -- (x1);
                        \draw [blue] (x1) -- (x0);
                        \draw [blue] (x0) -- (x2);
                        \draw [blue] (x2) -- (m2);
                                                \fill (m1) circle (1.5pt);
                        \fill (x0) circle (1.5pt);
                        \fill (x1) circle (1.5pt);
                        \fill (x2) circle (1.5pt);
                        \fill (m2) circle (1.5pt);
                                                \node (tau1) at (1,-0.2) {$\tau_1$};
                        \node (t1) at (2.5,-0.2) {$t_1$};
                        \node (t2) at (5,-0.2) {$t_1+t_2$};
                        \node (tau2) at (3.6,-0.2) {$\tau_2$};
                        \node (time) at (5.9,0) {time};
                        \node (space) at (0,3.2) {space};
                        \end{tikzpicture}}}\caption{Graphical representation of a contribution to the
                path-integral $Z^+ (m_1,t_1;x_0;m_2,t_2)$ given in Eq.~\eqref{PathIntegral}. The red curve represents the non-local interaction in the action (second line of Eq.~\eqref{ActionExpension}) while blue lines are bare propagators. There are two other  contributions when the time ordering is $\tau_1<\tau_2<t_1$ or $t_1<\tau_1<\tau_2$, already computed in \cite{WieseMajumdarRosso2010}.}
        \label{PathIntegralFigure}
\end{figure}

As $X_t$ is a Gaussian process, the action $S$ can  (at least formally) be constructed from the covariance function of $X_t$.  However this is not enough to evaluate the path integral \eqref{PathIntegral}  in all generality. Following the formalism of \cite{WieseMajumdarRosso2010}, we use standard Brownian motion as a starting point for a perturbative expansion, setting  $H = \frac{1}{2} + \varepsilon$ with $\varepsilon$ a small parameter; then  the action at first order in $\varepsilon$ is (we refer to the appendix of \cite{WieseMajumdarRosso2010} for the derivation)
\begin{equation}
\begin{split}
 S\left[ X \right] =&\; \frac{1}{4 D_{\varepsilon,\tau}} \int_0^t \dot{X}_{\tau_1}^2 \rmd\tau_1\\
& -\frac{\varepsilon}{2} \int_0^{t-\tau} \rmd\tau_1 \int_{\tau_1+\tau}^t \rmd\tau_2 \,\frac{ \dot{X}_{\tau_1} \dot{X}_{\tau_2}}{|\tau_2-\tau_1|}  +O(\varepsilon^2)\ .
\label{ActionExpension}
\end{split}
\end{equation}
The time $\tau$ is a regularization cutoff for coinciding times (one can also introduce discrete times  spaced by $\tau$ \cite{WieseMajumdarRosso2010}). The first line is the action for standard Brownian motion, with a rescaled diffusion constant \footnote{It is a dimensionfull constant as fBm and standard Brownian motion  do not  have the same {\em time dimension}.} $D_{\varepsilon,\tau} = 1 + 2 \varepsilon (1 + \ln(\tau)) + O(\varepsilon^2) \simeq (\rme\tau)^{2 \varepsilon} $. The second line is a correction,  non-local in time since  fBm is non-Markovian.

\begin{figure}[t]
	\Fig{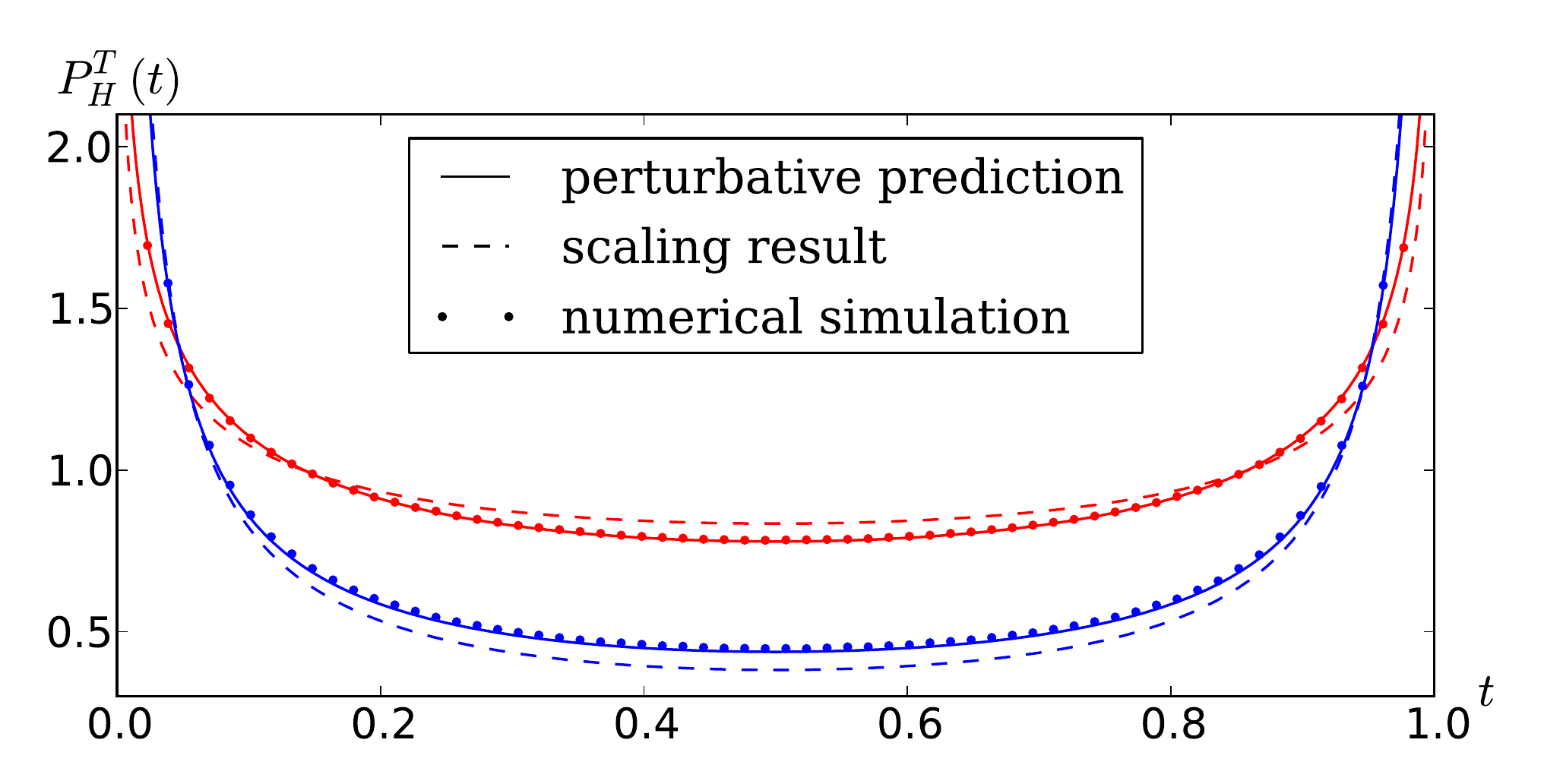}
	\caption{Distribution of $t_{\rm max}$ for $T=1$ and $H=0.25$ (red) or $H=0.75$ (blue) given in Eq.~\eqref{MaxPosDistrib} (plain lines) compared to the scaling ansatz, \textit{i.e.}\ $\mathcal{F}= \mbox{cst}$ (dashed lines) and  numerical simulations (dots). For $H<0.5$ realisations with $t_{\rm max}\approx T/2$  are less probable (by about $10\%$) than expected from  scaling. For $H>0.5$ the correction has the opposite sign.}\label{ArcSinLawFBM}
\end{figure}

Computing the $\varepsilon$ expansion of \eqref{PathIntegral} using \eqref{ActionExpension} is rather technical. A graphical representation of the key term is given in Fig.~\ref{PathIntegralFigure}.  The result for $Z^+(m_1,t_1;x_0;m_2,t_2)$ covers a page,   presented in \cite{DelormeWieseUnPublished}. 
We use this result here to deduce its most interesting implications, starting with the probability distribution of $t=t_{\rm max}$. For Browanian motion ($H=1/2$), this distribution is well known as the Arcsine law \footnote{The cumulative distribution of $t_{\rm max}$ involves the arcsin function.}, 
\begin{equation}
P^T_{\half}(t) = \frac{1}{\sqrt{\pi t(T-t)}} \ , \mbox{~for~} t\in [0,T] \ .
\end{equation}
Until now, only scaling properties were known for this distribution in the general case \cite{MajumdarRossoZoia2010b}.  The path integral \eqref{PathIntegral} is linked to this distribution via  
\begin{equation}
P^T_H(t) = \lim\limits_{x_0\rightarrow 0}\frac{1}{Z} \int_{m_1,m_2>0} Z^{+}(m_1,t;x_0;m_2,T-t)\ .
\label{TimeDistribDef}
\end{equation}
The normalization $Z$ depends  on $x_0$ and $T$.
Our result for
the distribution of $t_{\rm max}$ takes a nice forme if we exponentiate the order-$\varepsilon$ correction obtained from Eq.~\eqref{TimeDistribDef},
\begin{equation}
P^T_H(t)= \frac{1}{[t (T-t)]^H} \, \exp\! \left(\varepsilon\, \mathcal{F} \left(\frac{t}{T-t}\right)\right) +O\left(\varepsilon^2\right)\ .
\label{MaxPosDistrib}
\end{equation}
This is plotted on   Fig.~\ref{ArcSinLawFBM}.
We see the expected change in the scaling form of the Arcsine law, $\sqrt{t(T-t)} \rightarrow [t(T-t)]^H$ and a  non-trivial change in the shape given by the function
\begin{equation}
\begin{split}
\mathcal{F} (u)= &   \sqrt u \left[\pi-2 \arctan(  \sqrt u )\right]\\
&+\frac1 { \sqrt u }\left[\pi-2 \arctan\left(\frac1{ \sqrt u }\right)\right]+ \mbox{cst}\ .
\label{Fprediction}
\end{split}
\end{equation}
The time reversal symmetry $t\to T-t$  (corresponding to $u \rightarrow u^{-1}$) is explicit; the constant ensures normalization.

\begin{figure*}[t]
        \Fig{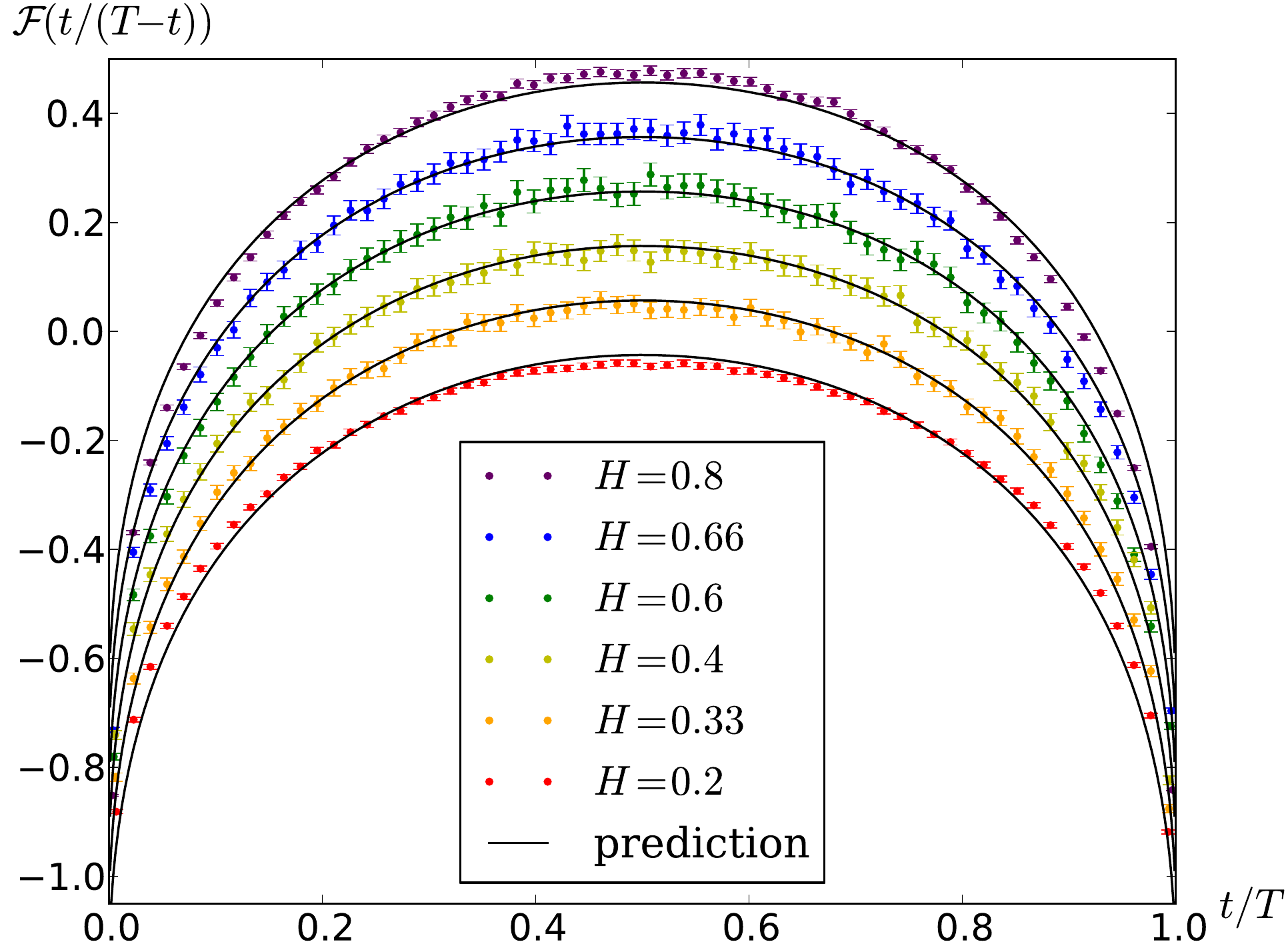}~~~\Fig{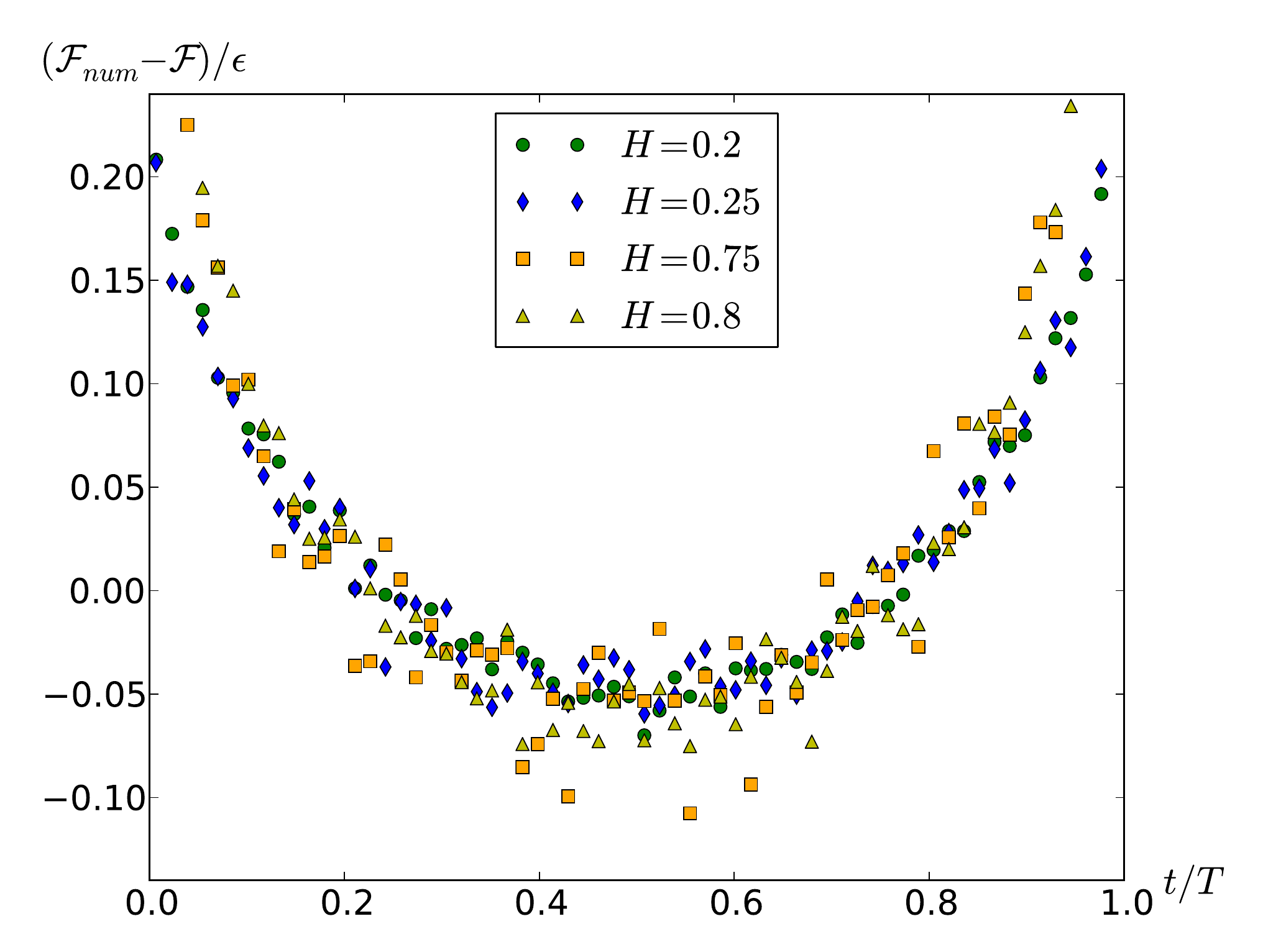}
        \caption{Left: Numerical estimation of $\mathcal{F}$ for different values of $H$ on a discrete system of size $N=2^{12}$, using $10^8$ realizations. Plain curves  represent  the  theoretical prediction \eqref{Fprediction},   vertically translated for  better visualization. Error bars are $2\sigma$ estimates. Note that for $H=0.6$, $H=0.66$  and $H=0.8$ the expansion parameter $\epsilon$ is positive, while for $H=0.4 $, $H=0.33$  and $H=0.2$ it is negative.
        Right: Deviation  for large $|\varepsilon|$ between the theoretical prediction \eqref{Fprediction} and the numerical estimations \eqref{Fnum}, rescaled by $\varepsilon$. These curves collapse for different  values of $H$, allowing for an estimate of the $O(\varepsilon^2)$ correction  to $P_H^T(t)$.}
        \label{ArcSinLawFBMnum}
\end{figure*}

We tested the prediction \eqref{MaxPosDistrib}-\eqref{Fprediction} with  numerical simulations of a discretized fractional Brownian motion for different values of $H$. To this aim, we used the  Davis and Harte  procedure as described in  \cite{Dieker} (and references therein). To compare numerical results with the theory, we extract an estimation $\mathcal{F}_{{\rm num}}^{\varepsilon}$ of the function $\mathcal{F}$\,as
\begin{equation}
\mathcal{F}_{{\rm num}}^{\varepsilon}\left(\frac{t}{T-t}\right):= \frac{1}{\varepsilon} \log \Big(P^{T,H}_{\rm num}(t) \times [t(T-t)]^H \Big)\ .
\label{Fnum}
\end{equation}
Here $P^{T,H}_{\rm num}(t)$ is the numerical estimation of the distribution of $t_{\rm max}$ for the discretized fBm at given $H$ (obtained with   uniform binning). Apart from discretization effects, we should see significant statistical errors as $\epsilon\to 0$, and systematic order-$\epsilon^2$ corrections for larger $\epsilon$. As can be seen on Figs.~\ref{ArcSinLawFBM} and \ref{ArcSinLawFBMnum}, our   numerical and analytical results are in  {\em remarkable} agreement for all values of $H$ studied, both for $\epsilon$ {\em positive} and {\em negative}.  As an example, for $H=0.75$, the correction to the pure scaling distribution has a relative magnitude of $10\%$ (see Fig.~\ref{ArcSinLawFBM}), which is measured in our simulation with a relative precision of $0.5\%$. This precision even allows  to numerically extract the subleading $O(\epsilon^2)$ correction, see Fig.~\ref{ArcSinLawFBMnum} right.

We now present results for the distribution of the maximum $P^T_H(m)$. For Brownian motion
\begin{equation}
P^{T}_{\half}(m) = \frac{e^{-\frac{m^2}{4T}}}{\sqrt{\pi T}}  \ , \qquad  m>0 \ . \label{MaxDistribSBM}\
\end{equation}
On the other hand, not much is known for generic values of $H$. This distribution is of  interest, as it is linked to  the survival probability $S(T,x)$, and the  persistence exponent $\theta$. The latter is defined for any random process $X_t$ with $X_0=x$ as\begin{equation}
\begin{split}
S(T,x)& = \mbox{proba}\left(X_t\geq 0 \mbox{ for all }t \in [0,T]\right)\\
& \underset{T \rightarrow \infty}{\sim} T^{-\theta_x}\ .
\end{split}
\end{equation}
For a large class of processes the exponent $\theta$ is independent of $x$. For fractional Brownian motion with Hurst exponent $H$ it was shown that $\theta_x=\theta = 1- H$ \cite{Molchan1999,Aurzada2011}. To understand the link of $S(T,x)$ with the maximum distribution for fBm, we use  self affinity of the process $X_t$ to write $P_H^T(m)$ as 
\begin{equation}\label{14}
P^T_H(m)=\frac{1}{\sqrt{2} T^H}f_H \!\left(\frac{m}{\sqrt{2} T^H}\right)\ .
\end{equation}
Here $f$ is a scaling function depending on $H$. Eq.~\eqref{MaxDistribSBM} can be  reformulated as $f_{\half}(y) = \sqrt{\frac{2}{\pi}} {e^{-y^2/2}}$. The survival probability is related to the maximum distribution by
\begin{equation}
S(T,x)=\int_0^xP^T(m)\,\rmd m = \int_0^{\frac{x}{ \sqrt{2} T^H}}f_H(u)\,\rmd u\ .
\end{equation}
This   states that a realisation of a fBm starting at $x$ and remaining positive is the same as a realisation starting at $0$ with a minimum larger than $-x$, due to  translation invariance of the fBm. Finally, the symmetry $x \rightarrow -x$ gives the correspondence between minimum and maximum.

\begin{figure*}[t]
        \Fig{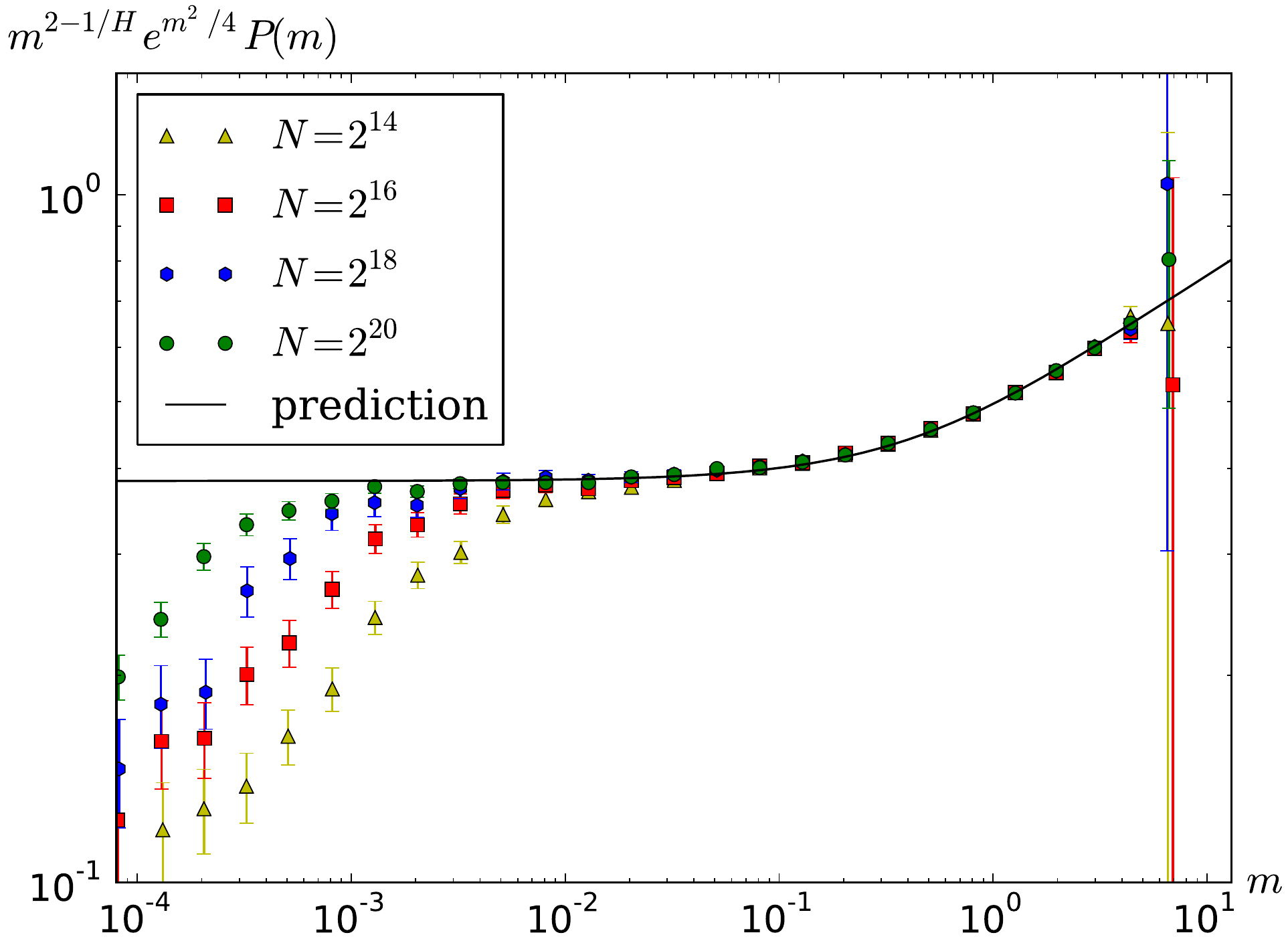}~~~\Fig{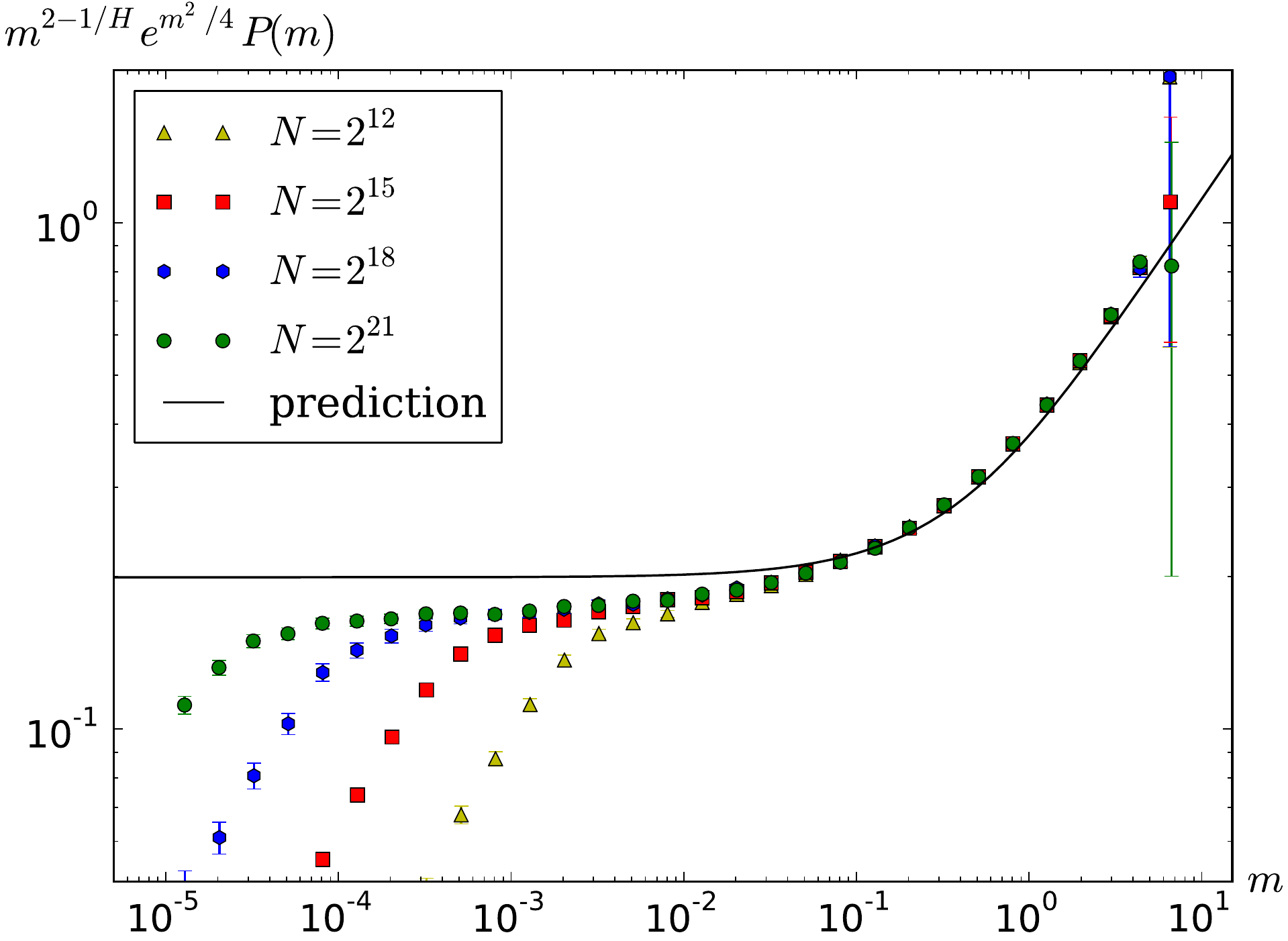}
        \caption{Left: The combination \eqref{MaxDistribTest} for $H=0.6$. The plain line is the analytical prediction $\exp({\varepsilon [\mathcal{G}(m/\sqrt{2}) + 4 \log m ]+\mbox{cst})}$ of the  distribution of the maximum without its small-scale power law and large-scale Gaussian behavior. The symbols are numerical estimations for $T=1$ of the same quantity $m^{2-{1}/{H}}\exp({ {m^2}/{4}})P^{ T=1,H}_{\rm num}(m)$ for different sample sizes. At small scale discretization errors appear.  At large scales the statistics is poor due to the Gaussian prefactor. For the  four decades in between theory and  numerics are in very good agreement. 
                Right: {\em ibid} for $H=0.75$.~The large scale-behavior on  both plots is consistent with  $m^{2\epsilon}$.                 
                }
        \label{MaxDistribNum}
                        
                \end{figure*}

These considerations allow  us to predict  the scaling behavior of $P^T_H(m)$ at small $m$ from the large-$T$ behaviour of $S(T,x)$ \cite{Molchan1999}, 
\begin{equation}
P^T_H(m) \underset{m \rightarrow 0}{\sim} m^{\frac{\theta}{H}-1}= m^{ \frac{1}{H}-2}\ .
\label{MaxScaling}
\end{equation}
Using our path integral, we can go further. The maximum distribution can be extracted from Eq.~\eqref{PathIntegral}, 
\begin{equation}
P^T_{H}(m) = \lim\limits_{x_0\rightarrow 0}\frac{1}{Z} \int_{0}^{T} {\rm d}t\int_{m_2>0} Z^{+}(m,t;x_0;m_2,T-t)\ .
\end{equation}
Its $\varepsilon$-expension leads to the scaling form of Eq.~(\ref{14}), with
\begin{eqnarray}\label{MaxDistribPrediction}
        f_{H}(y)= \sqrt{\frac{2}{\pi}}  e^{-\frac{y^2}{2}}  \,e^{\varepsilon \left[\mathcal{G}(y) + {\rm  cst}\right]} + O(\varepsilon^2)\ .
        \label{MaxDistrib}
\end{eqnarray}
The constant term ensures  normalization. The function $\mathcal{G}$  involves the hypergeometric function $_2F_2$:
\begin{equation}
\label{19}
\begin{split}
\mathcal{G}(y) =& \frac{y^4}{6}\, {}_2F_2\! \left(1,1; \frac{5}{2},3; \frac{y^2}{2} \right) - 3y^2\\
& + \pi (1-  y^2) \,\mathrm{erfi}\left( \frac{y}{\sqrt{2}} \right) + \sqrt{2 \pi} e^{\frac{y^2}{2}}y\\
&  + (y^2-2)\left[\gamma_{\rm E}+\ln \left(2y^2\right)\right]\ .
\end{split}
\end{equation}
This function has a different asymptotics for small and large $y$, 
\begin{equation}
\begin{split}
\mathcal{G}\left(y \right) \underset{y \rightarrow \infty}{\sim} &- 2\log(y) \\
\underset{y \rightarrow 0}{\sim} &-4 \log(y)\ .
\end{split}
\end{equation}
The second line implies that
$
P^T_H(m)  \underset{m \rightarrow 0}{\sim} m^{-4\varepsilon}
$
which is consistent (at order $\varepsilon$) with the scaling result  \eqref{MaxScaling},  $\frac{1}{H}-2= -4 \varepsilon + O(\varepsilon^2)$.
Formulas \eqref{MaxDistribPrediction}-(\ref{19}) also predict the  distribution at large $m$.   The leading behavior of $P^T_H(m)$ is Gaussian, which is well known, and can be derived from the {\em Borrel inequality} \cite{NourdinBook}. Our result for the  subleading term can be written as
\begin{equation}
\lim\limits_{y \rightarrow \infty} \frac{\log \left(f_H(y) \exp(\frac{y^2}{2}{ )} \right)}{\log(y)} = -2 \varepsilon + O(\varepsilon^2)  \ .
\label{asymp}
\end{equation}In order to test these predictions against numerical simulations, we can rewrite the form \eqref{MaxDistrib} s.t.\ the small-$m$ behavior matches the exact scaling result \eqref{MaxScaling}
\begin{equation}
f_H(y) = \sqrt{\frac{2}{\pi}} e^{-\frac{y^2}{2}}  y^{\frac{1}{H}-2} e^{\varepsilon \left[\mathcal{G}\left(y\right) + 4 \log y + {\rm  cst}\right]}  +O(\varepsilon^2) \ .
\label{MaxDistribPredictionCorrected}
\end{equation}
To extract the non-trivial contribution from numerical simulations, we study  for $T=1$ (see Fig.~\ref{MaxDistribNum})
\begin{equation}
m^{2-\frac{1}{H}}e^{ \frac{m^2}{4}}P^{{1,H}}_{\rm num}(m)= e^{\varepsilon \left[\mathcal{G}\left(\frac{m}{\sqrt{2}}\right) + 4 \log m + {\rm  cst} \right]} +O(\varepsilon^2)\ .
\label{MaxDistribTest}
\end{equation}
The sample size $N$ (\textit{i.e.}\ lattice spacing $ \rmd t=N^{-1}$) of the discretized fBm used for this numerical test is important, as $P_{\rm num}(m)$ recovers Brownian behavior for $m$ smaller than a cutoff of order $N^{-H}$. Far small   $H$ the necessary system size is very large, so we focus  on $H>0.5$. Figure \ref{MaxDistribNum} presents results for $H=0.6$ and $H=0.75$, without any fitting parameter. The constant term in the scaling form, relevant for normalization, is evaluated numerically. As   predicted,  convergence to the small-scale behavior is quite slow. This would lead to a wrong numerical estimation of the persistence exponent or other related quantities if the crossover to the large-scale behavior is not properly taken into account. At large scales, the  numerical data on Fig.~\ref{MaxDistribNum} grow as $m^{2\epsilon}$, consistent with the prediction~(\ref{asymp}). 

To conclude, we have given analytical results for the maximum of a fractional Brownian motion, and the time when this maximum is reached. To our knowledge these are the first analytical results for generic values of $H$ in the range $0<H<1$, beyond scaling relations.  Comparison to numerical simulations shows excellent agreement, even far from the expansion point $H=\half$.
 
 Our calculations also gave the joint probability of the maximum, the time when the maximum is reached, and the final point \cite{DelormeWieseUnPublished}. This allows us to address other observables of interests, such as fractional Brownian bridges.

\acknowledgements
We thank A.~Rosso for stimulating discussions, and PSL for support by grant ANR-10-IDEX-0001-02-PSL.


\end{document}